\documentclass[12pt,a4paper, headings = small]{article}

\usepackage[english]{babel}

\usepackage[ansinew]{inputenc}
\usepackage{graphicx} 
\usepackage{booktabs} 
\usepackage{color}    
\usepackage[lmargin = {2cm} , rmargin = {2cm} , tmargin = {2.5cm} , bmargin = {2.5cm}]{geometry}
\usepackage{titlesec} 
\usepackage{wrapfig}  
\usepackage{verbatim} 
\usepackage{amsmath}
\usepackage{amsthm}   
\usepackage{amssymb}  
\usepackage{textcomp} 
\usepackage{subfig}   
\usepackage{multirow} 
\usepackage{multicol}
\usepackage{array}
\usepackage{mathtools}
\usepackage{hyperref}
\usepackage{float}

\titleformat*{\section}{\large\bfseries}

\begin{document}

\title{The effect of the Austrian-German bidding zone split on unplanned cross-border flows}
\author{Theresa Graefe \thanks{Ulm University, Institute of Sustainable Management, 89081 Ulm, Germany, theresa.greafe@uni-ulm.de}}

\maketitle

\begin{abstract}
In 2013, TSOs from the Central European Region complained to the Agency for the Cooperation of Energy Regulators because of increasing unplanned flows that were presumed to be caused by a joint German-Austrian bidding zone in the European electricity market. This paper empirically analyses the effects of the split of this bidding zone in 2018 on planned and unplanned cross-border flows between Germany, Austria, Poland, the Czech Republic, Slovakia, and Hungary. For all bidding zones, apart from the German-Austrian one, planned flows increased. Further, I find that around the policy intervention between 2017 and 2019, unplanned flows between Germany and Austria as well as for the Czech Republic and Slovakia decreased. However, for Poland increasing unplanned flows are found.

\end{abstract}


\vspace{0.5cm}

\section{Introduction}
Until September 2018, Germany, Austria, and Luxembourg formed a joint bidding zone in the European power market. The countries thus traded electricity at a uniform price. Since much wind and solar power was generated at quasi-zero marginal costs in northern Germany, and traders were able to sell and buy as much electricity as they wanted, a huge amount of flows was scheduled from Germany to Austria without the need to allocate transmission capacities at this border (ACER, 2015). Already in 2013, a joint study of Transmission Systems Operators (TSOs) from the Central East European (CEE) region came to the conclusion that the electricity in many hours did not flow as it was traded.\footnote{The study was conducted by TSOs from Hungary (MAVIR), Poland (PSE), the Czech Republic (\v{C}EPS), and Slovakia (SEPS).} Instead, Poland, the Czech Republic, Slovakia, and Hungary had to host physical electricity flows that were traded between Germany and Austria (\v{C}EPS et al., 2013). Thus, in 2013, those TSOs complained to the Agency for the Cooperation of Energy Regulators (ACER) because of increasing unplanned flows that threatened grid stability in the host countries.\footnote{The Agency for the Cooperation of Energy Regulators (ACER) is a decentralized EU agency that ensures the integration of national energy markets and the implementation of legislation in the member states (ACER, 2021). It is located in Ljubljana, Slovenia.} 

\vspace{0.3cm}

In September 2018, Germany and Austria were eventually forced to split their joint bidding zone by the agency. Hence, two separate prize zones resulted: Austria and a German / Luxembourgian zone, further referred to as 'Germany', since Luxembourgian demand and trading activities can be seen as negligible. Through the split, the electricity trading process between Germany and Austria was changed. Along with the scheduling of flows, in separate auctions, transmission capacities now had to be bought. That means a restriction of trading opportunities for electricity traders. Politically, one goal of the split of the bidding zone was to reduce unplanned flows through the host countries (ACER, 2016). Thus, the primary objective of this paper is to analyze the impact of the split of the German-Austrian bidding zone on planned and absolute unplanned flows between Germany, Austria, Poland, the Czech Republic, Slovakia, and Hungary based on hourly data for two years around the splitting event. Apart from the unique event, the analysis of the joint bidding zone is interesting as the area was by far the largest consumer and trader in the European electricity market. In 2013, the bidding zone alone caused $65.5\%$ of the total yearly consumption in the CEE region (ACER, 2015). 

\vspace{0.3cm}
I find increasing planned flows for the complaining countries using a baseline OLS regression model. Between Germany and Austria, as expected, planned flows declined substantially. Regarding absolute unplanned flows, I find reductions for the German-Austrian border, as well as for the Czech Republic and Slovakia. For Hungary, absolute unplanned flows are not altered through the policy intervention. For Poland, absolute unplanned flows after the bidding zone split increase. The results are confirmed by the usage of a larger set of control variables and Post Lasso Double Selection and Partialling-out approaches, except for Poland, where the positive effect diminishes.\footnote{I use Post-Lasso Double Selection and Partiallin-out as implemented by Chernozhukov, Hansen, and Spindler (2019). They therefore provide the hdm package on R which can be found under the following link: https://cran.r-project.org/web/packages/hdm/index.html.} Also, further separating unplanned flows per border indicates that the situation improved for the Czech Republic concerning Germany and Austria, whereas unplanned flows between Poland and Germany and the Czech Republic rather increased. Therefore, I conclude that the policy intervention did indeed reach the intended goal of reducing unplanned flows for Slovakia and most of the Czech Republic. For Hungary, no effects are found and for Poland, it seems like the policy intervention was leading to increasing absolute unplanned flows. Still, for Poland, the situation remains the least clear.
\vspace{0.3cm}

The remainder of this paper is structured as follows: In section 2, a summary of the related literature is provided. Section 3 describes the institutional setup of European cross-border electricity trading. Section 4 demonstrates the hypothesis and the estimation procedure, before in section 5 the used data and some descriptive statistics are presented. In section 6, the main empirical results are shown and discussed. In section 7, Post Lasso Double Selection and Partialling-out are performed as robustness checks, which confirm the main results. The paper ends with some concluding remarks.

\vspace{0.3cm}

\section{Related Literature}

So far, the literature covering cross-border electricity flows is rather sparse. \v{C}EPS et al. (2013) descriptively quantify unplanned flows based on historical values for the CEE region. Taniguchi (2020) compares the role of renewable electricity generation systems in Denmark and Japan on commercially scheduled interconnector flows with a multiple regression model. Further, Sk\r{a}nlund  et al. (2013) use a correlation analysis of the central European power system to assess unplanned flows connected to the German-Austrian bidding zone.

\vspace{0.3cm}

Moreover, several drivers of electricity flows related to the German-Austrian bidding zone are identified in the literature: By using principal component analysis and local polynomial regression, Zugno, Pinson, and Madsen (2013) find that wind power generation especially from Germany has a substantial effect on physical cross-border flows in the European market. In line with that, J\'{o}nsson, Pinson, and Madsen (2010) describe that the increase in renewable electricity production in northern Germany pushes through the grids of neighboring countries. Sk\r{a}nlund  et al. (2013) find correlations between wind feed-in in northern Germany as well as scheduled flows between Austria and Germany with unscheduled flows between Germany and Poland. Signh et al. (2016) investigate unplanned flows between Germany, Austria, Poland, and the Czech Republic with an optimal power flow simulation based on data from the year 2013. They conclude that unplanned flows through Poland and the Czech Republic are mainly caused by northern German wind generation in combination with a poor transmission infrastructure at the German-Austrian interconnector.  ACER (2019) adds the absence of sufficient coordination of the capacities, and APG (2021) argues that lagging transfer grid development consequently reinforced unplanned electricity flows. In 2017, after the decision of the ACER to split the joint bidding zone, the negatively affected TSOs made a press release public. In this, they describe that instead of coordinated capacity allocation, as demanded by the ACER, a bilateral agreement between Germany and Austria about the congestion management at their border was taken. The TSOs thus raise the concern that the split of the bidding zone alone may not lead to an improved situation as without coordinated capacity allocation there is still a preference for energy traded between Germany and Austria (\v{C}EPS et al., 2017).

\vspace{0.3cm}

Some other papers cope with the topic of bidding zone reconfiguration in a more general way. Bem\v{s}, Kr\'{a}l\'{i}k, and Kn\'{a}pek (2016) qualitatively describe, which criteria should be applied in examining the effectiveness of bidding zones. The authors mention congestion rents, remedial actions, price differentials and volatility and transition costs, as well as social welfare. Borowski (2020) compares the traditional electricity system that is based on bidding zones with a nodal model with local electricity prices at each node. He argues that nodal pricing sets better price signals and fosters investment in the electricity market. Another strand of literature covers the integration of markets in the European Union: Antweiler (2016) shows that scheduled electricity exports and imports are mainly driven by price differences between two bidding areas. Gugler, Haxhimusa, and Liebensteiner (2018) and Keppler, Phan, and Le Pen (2017) studied the integration of European electricity prices and found that increasing variable renewable electricity production for example in northern Germany disturbs market integration, leading to a higher frequency of unplanned flows, congestion occurrences, and higher price differentials. Keppler, Phan, and Le Pen (2017) thereby show that market coupling causes converging prices between France and Germany but that this effect is counteracted by a high amount of renewable intermittent electricity production in Germany. To the best of my knowledge, so far, there is no analysis of the effects of bidding zone splits and no approach that uses unplanned flows as a dependent variable in an empirical analysis.

\vspace{0.3cm}

\section{Cross-border trading in the European electricity market}

When power is traded at the electricity exchange, the outcome is referred to as commercially traded or planned flows. Planned flows are day-ahead commercial exchanges based on matched nominations from previous time horizons (ENTSO-E Transparency Platform, 2021). The real movement through the grid is represented by physical flows. Planned and physical flows can deviate, as electricity does not necessarily flow as it is traded. Instead, electricity follows the path of least resistance, which depends on the available transfer capacities in the grid. Until September 2018, the joint Austrian-German bidding zone existed. Figure 1 shows pre-treatment averages of hourly planned and physical flows in the observed region, measured in GWh.\footnote{The pre-treatment sample reaches from September 2017 until the policy intervention in September 2018.} The averages are calculated as the sum of physical and planned flow in both directions, e. g. from Germany to Austria and from Austria to Germany. The observed sample in this study covers cross-border flows between Germany, Austria, Poland, the Czech Republic, Slovakia, and Hungary. 

\vspace{0.3cm}

Figure 1 shows that the deviation in averages of pre-treatment physical and planned flows is small between Hungary and Austria and between Slovakia and Hungary. \v{C}EPS et al. (2013) already discussed that unplanned flows at the Hungarian and Slovakian borders are within an acceptable range. Instead, they identify Poland and the Czech Republic as the two countries that are mainly affected by unplanned transactions related to the joint bidding zone. Descriptively, the pre-treatment averages seem to confirm this. The largest deviations can be found at the German-Austrian border. Here, average planned flows ($6.16$ GWh) largely exceed average physical flows ($1.52$ GWh). The same, but to a lesser extent, is true for the connections Germany-Czech Republic and Slovakia-Czech Republic. For the rest, physical flows are on average larger than planned flows (DE-PL, AT-CZ, PL-CZ, PL-SK). Hence, before the bidding zone split, in many hours power was scheduled from lower-priced northern Germany to demand centers that were located in the South of Germany and Austria.

\begin{figure}[h!]
\centering
\caption{Average pre-treatment planned and physical cross-border flows}
	\includegraphics[width=0.9\textwidth]{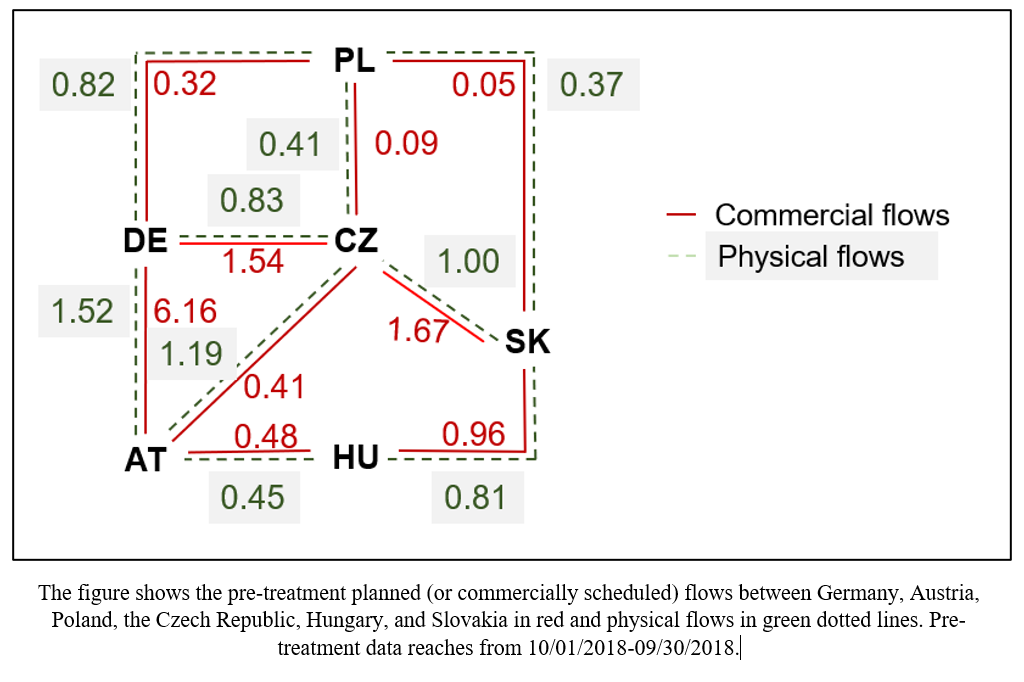}

\end{figure}

 As intra-zonal available transfer capacities often were not sufficient, physical flows are presumed to follow a path through eastern Europe, especially through Poland and the Czech Republic.\footnote{By intra-zonal capacities, transmission capacities between Germany and Austria are meant as well as transmission capacities within Germany.} Since those flows were not planned by the host countries, they are called unscheduled or unplanned flows (ACER, 2018). Those unplanned flows are calculated as the difference between physical flows and commercially traded flows in hour \textit{t} (ENTSO-E, 2014). The index \textit{i} here either stands for the border (Germany-Austria) or the country (the Czech Republic, Poland, Slovakia, or Hungary):

\begin{equation} \label{eq:sample2}
\begin{array}{ll}
unplanned_{i,t} =  physical_{i,t} - planned_{i,t}
\end{array}
\end{equation}

\vspace{0.2cm}

Unplanned flows are problematic if the source and the sink of those transactions are in the same bidding zone.\footnote{ACER (2018) call those kind of flows loop flows, as they stem from bidding zone internal exchanges but travel through neighboring bidding zones.} In that case, unplanned flows do not impact the net exchange position of neighboring countries which means, that the hosts do not get compensated for accommodating those unplanned physical movements through their grid (Bem\v{s}, Kr\'{a}l\'{i}k, and Kn\'{a}pek, 2016). ACER (2015) states that between 2011 and 2014, around $48\%$ of the flows that were scheduled between Germany and Austria were flowing through neighboring countries. For the years 2011 and 2012, \v{C}EPS et al. (2013) find that about $50\%$ of the scheduled flows within the German-Austrian bidding zone were adversely affecting Poland and the Czech Republic. To reduce structural congestion in the host countries, ACER (2015) suggested the split-up of the German-Austrian bidding zone and the introduction of coordinated capacity allocation at the German-Austrian border which was eventually conducted in September 2018. With the separate bidding areas, traders are only able to buy and sell as much electricity between Germany and Austria, as transmission capacities are available. The capacities at this border are limited to $4,900$ MWh  (Bundesnetzagentur, 2021).

\vspace{0.3cm}

There are several reasons for the decision: Unplanned flows afflict external costs to the host countries, as they have to be incorporated into the grid (ACER, 2015). The capacities that would be available for commercial cross-border trading may have to be reduced by the affected TSOs. So, the host countries cannot use their transmission capacities in other, more lucrative transactions (Bundesnetzagentur, 2021). This means that the TSOs lose benefits compared to a situation with sufficient transfer capacities. According to ACER (2016), this is reinforced due to unequal treatment of electricity exchanges of flows between and within bidding areas: The handling of internal exchanges within a bidding zone is discriminatory concerning cross-zonal exchanges between two different bidding zones. While exchanges between different bidding areas are constrained by the TSO through the allocation of transfer capacities, within a bidding zone, consumers and generators contract electricity in an unconstrained manner. Also, the network is forced to accommodate all physical flows that stem from intra-zonal exchanges whereas cross-zonal exchanges have disadvantages in network access (ACER, 2016). This induces costs of reduced capacities for trading in the host countries and a loss of congestion revenues for the TSOs. As well, the affected countries face costs related to the security of supply through expensive remedial actions. For example, the host countries have to reduce their production to ensure grid stability. This inflicts costs for re-dispatching, counter trading, and curtailment (ACER, 2015). According to Sk\r{a}nlund et al. (2013), those re-dispatch actions are expensive and negatively affect network security. ACER (2018) further states that the force to offer less cross-zonal capacity for trade due to unequal treatment of electricity exchanges reduces the efficiency of the market. Moreover, wrong signals are created that do not foster investment in the development of the transmission system (ACER, 2015).

\section{Hypothesis and Empirical Strategy}

In this section, I describe the econometric estimation procedure to investigate, how the bidding zone split affected planned and unplanned cross-border flows in the observed region. I here follow the argumentation of the complaining TSOs from the CEE region and the ACER, that unplanned flows before the split are related to the joint German-Austrian bidding zone. Therefore, several hypotheses are to be tested. First of all, I use a baseline specification that is estimated by Ordinary Least Squares estimation. The dependent variable \textit{$planned_{i,t}$} is planned flows for the border, which is the case for flows between Germany and Austria or a country \textit{$i=1,2,...n$} in hour \textit{$t =1,2,...,T$}:

\begin{align*}
planned_{i,t} = \alpha_0 + \alpha_1 \delta_{t} +  \alpha_2 {X}_{i,t} +  \tau_t + \eta_{i,t}\qquad \qquad  (2)
\end{align*}

Further, $\delta_t$ is a dummy variable that indicates whether an observation is pre- (0) or post-treatment (1). Thus, $\hat{\alpha}_1$ is the estimate for the parameter of interest. Moreover, $X_{i,t}$ is a vector of 107 control variables. For the four German TSOs 50Hertz, Tennet, TransnetBW, and Amprion and the Austrian APG, forecasted onshore and offshore wind generation data is collected. As well, on that spatial level, forecasted solar PV generation and residual load values are part of $X_{i,t}$. Aggregated at the country level, electricity generation from other sources (e. g. lignite, biomass, ...) for Germany is included. Renewable and fossil electricity generation for Austria, the Czech Republic, Poland, Slovakia, and Hungary is added along with loads and residual load differences between the countries. Solar PV and wind generation are further allowed to take a quadratic form. As the chosen variables might be correlated over time, affecting the dependent variable, they have to be included. If they vary over time, they otherwise would bias the estimate of interest if not controlled for. Further, I assume all control variables to be exogenous.\footnote{Electricity generation can be seen as exogenous source of variation of (un-)planned cross-border flows. According to Cl\'{o}, Cataldi, and Zoppoli (2015), short-term electricity demand can be seen as exogenous, too.} With $\tau_t$, fixed effects for the hour of the day, the month of the year, and the weekday are added. $\eta_{i,t}$ is an error term. I form the following hypotheses:

\vspace{0.3cm}

\textbf{Hypothesis 1}: \textit{After the policy intervention, planned flows at the German-Austrian border will decrease.}

\vspace{0.1cm}

\textbf{Hypothesis 2}: \textit{After the policy intervention, planned flows for Poland and the Czech Republic will increase.}

Hypotheses 1 and 2 reflect the expectations of the behavior of commercially traded (or planned) flows following the policy intervention. According to Bundesnetzagentur (2021), with the split of the joint bidding zone, the transmission capacities (and thus commercially scheduled flows) between Austria and Germany are limited to $4,900$ MW. As there has been no limitation of commercially traded flows before the split, those planned flows are expected to go down. For the estimation, planned flows from Germany to Austria and in the opposite direction are aggregated. As cheap electricity is still produced in Northern Germany after the policy intervention and traders will still be interested in trading the electricity at a low price, I expect, that trading activities will remain at a high level. Since the policy intervention forces traders to reduce the scheduled flows between Germany and Austria, I expect to find increasing planned flows for the complaining countries after the policy intervention. To observe the effect of the bidding zone split, cross-border trading activities are aggregated at the country level.\footnote{For the Czech bidding zone (CZ), I sum up the following planned flows: CZ-DE, DE-CZ, PL-CZ, CZ-PL, CZ-AT, AT-CZ, CZ-SK, SK-CZ. For the Polish bidding zone (PL) flows from PL-DE, DE-PL, CZ-PL, PL-CZ, SK-PL, and PL-SK are aggregated. For Slovakia (SK) flows from SK-PL, PL-SK, CZ-SK, SK-CZ, HU-SK, and SK-HU, and for Hungary (HU) AT-HU, HU-AT, SK-HU, HU-SK are summed up. Although Austria and Slovakia share a border, no electricity trading data is found for the interconnection between the two countries.} 

\vspace{0.3cm}

Secondly, I define absolute unplanned flows \textit{$|unplanned_{i,t}|$} as the dependent variable:
\begin{align*}
|unplanned_{i,t}| = \beta_0 + \beta_1 \delta_{t}+  \beta_2 {Z}_{i,t} +  \tau_t + \varepsilon_{i,t}\qquad \qquad  (3)
\end{align*}

As described, unplanned flows are calculated as the difference between physical and planned flows. Every deviance of physical from planned flows is probable to cause difficulties for the countries that have to cope with them. As I am mainly interested in problems for the providers and all unplanned flows are problematic, I take absolute values for the analysis. $Z_{i,t}$ again is a vector of 107 control variables. Instead of forecasted renewable generation values from the German and Austrian TSOs, here the difference between forecasted and actual generation values are taken as controls. Further, if absolute unplanned flows for Poland, the Czech Republic, Slovakia, or Hungary are the dependent variable, scheduled flows from Germany to Austria and in the opposite direction are added as additional control variables, as in the literature they are identified as a driver of unplanned flows in the region. I form two more hypotheses:

\vspace{0.3cm}

\textbf{Hypothesis 3}: \textit{After the policy intervention, absolute unplanned flows at the German-Austrian border will go down.}

As \v{C}EPS et al. (2013) already noticed for 2011 and 2012, and Figure 1 shows for the pre-treatment period, planned flows between Germany and Austria massively exceed physical flows at the German-Austrian. If planned flows at the border indeed decreased and physical flows stayed the same after the policy intervention, I expect the negative unplanned flows to become less negative. In absolute terms, this means decreasing unplanned flows. 

\vspace{0.3cm}

\textbf{Hypothesis 4}: \textit{After the policy intervention, absolute unplanned flows for Poland and the Czech Republic will go down.}

The split of the joint bidding zone was targeted at unplanned flows in the region. If the policy intervention was successful, absolute unplanned cross-border flows for the negatively affected countries are expected to decrease. This is the main hypothesis to test the success of the policy intervention. 

\vspace{0.3cm}

\section{Data and Descriptive Statistics}

All data used in this study are publicly available and joined at hourly resolution for Germany, Austria, Poland, the Czech Republic, Hungary, and Slovakia. The sample reaches from 01st October 2017 until 30th September 2019, covering a balanced two-year period around the policy intervention. The data frame thus comprises 17,518 observations.\footnote{As there are only a few missing values and the sample is large, missing values are replaced by column means.} The policy intervention took place at midnight between 30th September and 1st October 2018. That results in 8,759 data points before the split of the German-Austrian bidding zone and 8,759 afterwards. Physical and commercial trade data are taken from ENTSO-E Transparency Platform, as well as electricity generation data for all countries. Additionally, more detailed actual and forecasted renewable generation values for the four German TSOs and APG are taken from the Bundesnetzagentur. Also, residual load values for the TSOs stem from this source. Unplanned flows and residual load differences are calculated manually. Unplanned flows are computed as described in equation (1).\footnote{For instance, absolute unplanned flows between Germany and Austria are calculated as $|physical flows_{DE\_AT} - planned flows_{DE\_AT}| + |physical flows_{AT\_DE} - planned flows_{AT\_DE}|$.} For the residual load differences, residual loads first have to be obtained. For Austria, values are taken from Bundesnetzagentur. For Germany, residual load values of the four German TSOs are aggregated. For Poland, the Czech Republic, Slovakia, and Hungary forecasted load values are taken from ENTSO-E Transparency Platform and wind and solar PV generation is subtracted. For the six countries, then residual load differences are calculated. All data are given in Gigawatt-hours (GWh).\footnote{Data from ENTSO-E Transparency Platform are accessed in 12/2020 (https://transparency.entsoe.eu/); data from Bundesnetzagentur are accessed in 01/2021 (https://www.smard.de/home/downloadcenter/download-marktdaten).} 

\vspace{0.3cm}

Table 1 depicts the averages and standard deviations of planned cross-border flows, measured in GWh, before and after the policy intervention.

\begin{table}[h!]
	\begin{center}
		\begin{tabular}{  m{2cm}   m{2cm}  m{2cm}   m{2cm}  m{2cm}  m{2cm}  m{2cm} } 
			\hline
			\hline
			&  \multicolumn{2}{l}{\textbf{Before split}}  &     \multicolumn{2}{l}{\textbf{After split}}  & \textbf{Difference} \\ 
			
			& \textbf{Mean} & \textbf{S. D.} & \textbf{Mean} & \textbf{S. D.} & \\
			\hline
			\multicolumn{3}{l}{Planned flows (GWh):}&  \\
			\hline
			DE-AT & 6.16 &	1.05	& 	2.81 &	1.53	& 	-3.35 \\
			\hline
			CZ &  3.71  &  0.77 &	 4.08	 &   0.85 &	 0.37 \\
			\hline
			PL & 0.46 & 	0.33 &  0.56 &	 0.37 &	0.1 \\
			\hline
			SK & 2.68 & 0.62 & 2.74 & 0.7 & 0.06 \\
			\hline
			HU & 1.44 & 0.37 & 1.69 & 0.36 & 0.25 \\
			\hline
			\hline

		\end{tabular}
	\end{center}
	\caption{Descriptive statistics of planned cross-border flows}
\end{table}

 The data are aggregated for the German-Austrian border, Poland, the Czech Republic, Slovakia, and Hungary as described in section 4. The table shows that descriptively planned flows massively decreased for the German-Austrian interconnection after the split of the bidding zone. The difference in the averages amounts to $-3.35$ GWh. For the four other bidding zones, average planned cross-border flows increased. Descriptively, the largest increase is shown for the Czech Republic with $0.37$ GWh, followed by Hungary with $0.25$ GWh. For Poland and Slovakia, the increases are quite small with $100$ and $60$ MWh. Apart from that, for all units, except for Hungary, the standard deviation after the policy intervention goes up compared to before.

\vspace{0.3cm}

Table 2 descriptively compares pre- and post-treatment averages of absolute unplanned flows. One can see, that descriptively absolute unplanned flows decreased for the German-Austrian border, in line with hypothesis 3. The same applies to Slovakia. Regarding the Czech Republic, and Hungary the differences are close to zero. For Poland, the comparison shows a slight increase.\footnote{Figure 3 in Appendix A plots unplanned cross-border flows (not absolute values).  It shows that indeed unplanned flows between Germany and Austria are utterly negative before the split of the bidding zone. After the policy intervention, values are closer to zero and often within a positive range. Plotting unplanned flow for the Czech Republic instead of absolute values shows  that also here, values are often negative but rather distributed around zero. It looks like unplanned flows after the split vary more, especially in the negative direction. For Poland, the development looks quite similar but after the intervention unplanned flows seem to be a little more positive. For Slovakia and Hungary, unplanned flows seem to vary less and negative values appear to become less frequent after the bidding zone split.} 
\vspace{0.3cm}

\begin{table}[h!]
	\begin{center}
		\begin{tabular}{  m{2cm}   m{2cm}  m{2cm}   m{2cm}  m{2cm}  m{2cm} m{2cm} } 
			\hline
			\hline
			&  \multicolumn{2}{l}{\textbf{Before split}}  &     \multicolumn{2}{l}{\textbf{After split}}  & \textbf{Difference} \\ 
			
			& \textbf{Mean} & \textbf{S. D.} & \textbf{Mean} & \textbf{S. D.} & \\
			\hline
			\multicolumn{3}{l}{Absolute unplanned flows (GWh):}&  \\
			\hline
			DE-AT & 4.64  &1.4 & 1.45 & 0.93 &  -3.19  \\
			\hline
			CZ &  1.03 & 0.79 & 1.09  & 0.8 & 0.06   \\
			\hline
			PL & 1.15 & 0.6 &  1.29 & 0.67 &  0.14 \\
			\hline
			SK & 0.7 & 0.65 & 0.49 & 0.44 &  -0.21 \\
			\hline
			HU & 0.34 & 0.31 & 0.3 & 0.24 & -0.04 \\
			\hline
			\hline
		\end{tabular}
	\end{center}
	\caption{Descriptive statistics of absolute unplanned flows}
\end{table}

\newpage

\section{Results}

I estimate equation (2) by Ordinary Least Squares estimation to test hypotheses 1 and 2 for the data set. Table 3 presents the results for the estimator of the parameter of interest $\hat{\alpha}_1$ with planned flows between Germany and Austria (DE-AT), for the Czech Republic (CZ) and Poland (PL) (and additionally for Slovakia (SK), and Hungary (HU)) as the dependent variables. I find estimators that are significant at the $1\%$ level for the joint German-Austrian bidding zone, Poland, and Hungary.\footnote{Heteroskedasticity and autocorrelation robust standard errors after the Newey-West (1994) method in parentheses.} Further, all coefficients of the zone splitting dummy $\delta$ receive estimates in the expected directions. As formulated in hypothesis 1, the effect of the bidding zone split on planned flows between Germany and Austria is strongly and significantly negative. The reduction amounts to $2.914$ GWh.

\begin{table}[h!]
\begin{center}
\scalebox{0.8}{
\begin{tabular}{ m{3cm} m{3cm} m{3cm} m{3cm} m{3cm} m{3cm}} 
\hline
\hline
  \multicolumn{6}{c}{\textbf{Dependent variable: Planned flows} } \\ 
\hline \\[-1.8ex] 
 & DE-AT & CZ & PL & SK & HU\\ 
& (1) & (2) & (3) & (4) & (5) \\ 
\hline \\[-1.8ex] 
delta (zone splitting dummy) & $-$2.914$^{***}$ & 0.139  & 0.350$^{***}$ & 0.073  & 0.281$^{***}$ \\ 
  & (0.230) & (0.173) & (0.056) & (0.119) & (0.060)  \\ 
  &   \\  
\hline \\
Controls & Yes & Yes & Yes & Yes & Yes \\
Fixed Effects & Yes & Yes & Yes & Yes & Yes \\
Observations & 17,518  & 17,518  & 17,518  & 17,518  & 17,518  \\ 
Adjusted R$^{2}$ & 0.811 & 0.305 & 0.353 & 0.403 & 0.452 \\ 
\hline
\textit{Note:}  & \multicolumn{5}{r}{$^{*}$p$<$0.1; $^{**}$p$<$0.05; $^{***}$p$<$0.01} \\ 
 \multicolumn{6}{l}{Standard errors are HAC robust. Lags are chosen with the Newey-West (1994) method.} \\
 \hline
 \hline
\end{tabular}}
\end{center}
\caption{Regression results planned flows}
\end{table}

 On average planned flows between Austria and Germany are $2.914$ GWh lower in post-treatment hours compared to pre-treatment hours. With respect to pre-treatment average, this is equal to a reduction of $47\%$. As well, for the four observed countries, planned flows increased after the policy intervention. The largest and significant effect arises for Poland, where post-treatment planned flows on average increased by $0.35$ GWh (or by $+76\%$) compared to pre-treatment hours. This effect is much larger than the sole comparison of pre- and post-treatment averages as is done in Table 1. Also, for Hungary, the effect is strong with $0.281$ GWh (respectively $+11\%$) and nearly equals the simple time differences. For the Czech Republic and Slovakia the estimated treatment effects are as well positive but not significant. The Czech increase in relative terms is quite small with only $3\%$. But here, already before the treatment, planned flows were around eight times as high as in Poland, so the starting point for that country is already at a high level. Still, the estimates for the split coefficients are well away from zero. Accordingly, hypotheses 1 and 2 cannot be rejected. Table 7 in Appendix B further shows the impact of the bidding zone split on physical flows.

 \vspace{0.3cm}
 To answer the question of how the bidding zone split influenced absolute unplanned flows, moreover equation (3) is estimated by OLS. Table 4 presents the regression results. Hypothesis 3 reflects the expectation that absolute unplanned flows between Germany and Austria declined through the policy intervention. Indeed, I find a strong negative and significant effect of  $-2.696$ GWh for that border, so hypothesis 3 cannot be rejected.

\vspace{0.3cm}

\begin{table}[h!]
\begin{center}
\scalebox{0.8}{
\begin{tabular}{ m{3cm} m{3cm} m{3cm} m{3cm} m{3cm} m{3cm}} 
\hline
\hline
  \multicolumn{6}{c}{\textbf{Dependent variable: Absolute unplanned flows}} \\ 
\hline \\[-1.8ex] 
 & DE-AT & CZ & PL & SK & HU\\ 
 & (1) & (2) & (3) & (4) & (5)\\ 
\hline \\[-1.8ex] 
delta (zone splitting dummy) & $-$2.969$^{***}$ & $-$0.408$^{***}$  & 0.312$^{**}$ & $-$0.314$^{***}$ & 0.025\\ 
  & (0.245) & (0.147) & (0.132) &  (0.092) & (0.052) \\ 
  &  \\  
\hline \\
Controls  & Yes & Yes & Yes & Yes & Yes\\
Fixed Effects  & Yes & Yes & Yes & Yes & Yes \\
Observations & 17,518 & 17,518 & 17,518 & 17,518 & 17,518 \\
Adjusted R$^{2}$ & 0.790 & 0.330 & 0.517 & 0.388 & 0.306 \\ 
\hline
\textit{Note:}  & \multicolumn{5}{r}{$^{*}$p$<$0.1; $^{**}$p$<$0.05; $^{***}$p$<$0.01} \\ 
 \multicolumn{6}{l}{Standard errors are HAC robust. Lags are chosen with the Newey-West (1994) method.} \\
\hline
 \hline
\end{tabular}}
\end{center}
\caption{Regression results absolute unplanned flows}
\end{table}

Compared to the pre-treatment average, the reduction is equal to $64\%$. The complaining countries that claimed to suffer the most from unplanned flows in the CEE region were Poland and the Czech Republic. If the policy intervention had the intended effect, I expect absolute unplanned flows to decrease for the two countries, as stated in hypothesis 4. According to the regression results, hypothesis 4 can be significantly confirmed at the $1\%$ level for the Czech Republic. On average, absolute unplanned flows are $408$ MWh ($-39.6\%$) lower in post-treatment hours compared to pre-treatment hours. The results are also in line with those from the planned and physical flows regressions. Before the split, on average one observed negative unplanned flows. As shown, planned flows increased slightly more than physical flows which should result in an overall decrease in (absolute) unplanned flows.\footnote{Also, for Slovakia, I find a significant reduction of $314$ MWh, which means a decline of $44.9\%$ compared to pre-treatment averages.For Hungary, the effect is only small and insignificant ($25$ MWh or $7.4\%$).} What is not as expected, is that for Poland, the effect is strong and significantly positive (significant at the $5\%$ level). Here, on average absolute unplanned flows increased by $312$ MWh after the policy intervention. Compared to the pre-treatment average of $1.150$ MWh this is equal to an increase of $27.1\%$. Hence, the results suggest, that hypothesis 4 is rejected for Poland. \footnote{Interestingly, this is not in line with the former regressions. In Poland, unplanned flows before the policy intervention were positive ($1.6$ GWh), meaning that physical flows were higher than planned flows. Through the policy intervention, planned flows increased more than physical flows ($0.35$ GWh compared to $0.28$ GWh). Thus, the difference between physical and planned flows should become smaller, which should lead to decreasing (absolute)unplanned flows.Still, Table 4 does not indicate that a relief through the bidding zone split is apparent.}
 \vspace{0.3cm}

To better understand where the unplanned flows exactly occur, I re-estimate equation (3) but with separated values per border. The results are shown in Table 5.

\begin{table}[h!]
\begin{center}
\scalebox{0.8}{
\begin{tabular}{ m{3cm} m{3cm} m{3cm} m{3cm}  m{3cm} } 
\hline
\hline
\multicolumn{5}{c}{\textbf{Dependent variable: Absolute unplanned flows separated per border}} \\ 
\hline \\[-1.8ex] 
 & CZ-DE &  CZ-PL & CZ-AT & CZ-SK   \\ 
\hline \\[-1.8ex] 
 delta & $-$0.078 &  0.201$^{***}$ & $-$0.234$^{***}$   & $-$0.297$^{***}$ \\ 
  & (0.092) & (0.054) &  (0.069)  & (0.080)   \\ 
&  \\
\hline \\
Adjusted R$^{2}$ & 0.225 & 0.419 & 0.741 & 0.424 \\ 
\hline
  & \\
& PL-DE &  PL-SK &  SK-HU  & HU-AT \\
\hline \\[-1.8ex] 
delta & 0.156$^{**}$ & $-$0.045 &  0.028 & 0.040  \\
 & (0.071)&  (0.036) & (0.032) & (0.030)\\
 &  \\
\hline \\
Adjusted R$^{2}$ & 0.502 & 0.498 & 0.147 & 0.225 \\ 
\hline
\textit{Note:}  & \multicolumn{4}{r}{$^{*}$p$<$0.1; $^{**}$p$<$0.05; $^{***}$p$<$0.01} \\ 
 \multicolumn{5}{l}{Standard errors are HAC robust. Lags are chosen with the Newey-West (1994) method.} \\
\hline
 \hline
\end{tabular}}
\end{center}
\caption{Regression results absolute unplanned flows per border}
\end{table}

 CZ-DE stands for the sum of absolute unplanned flows from the Czech Republic to Germany and in the opposite direction, from Germany to the Czech Republic. Observing the borders in the region separately, yields a more diverse picture. For the Czech Republic, reductions in unplanned flows stem from interconnections with Germany ($-78$ MWh), and mainly from the interconnections with Austria ($-234$ MWh) and Slovakia ($-297$ MWh). Still, absolute unplanned flows increased significantly between the Czech Republic and Poland by $201$ MWh. As well, unplanned flows between Poland and Germany grew by $156$ MWh. Results for PL-SK, SK-HU, and HU-AT are small and insignificant. Hence, also the analysis of the directed flows does indicate that the situation improved for Slovakia and at large also for the Czech Republic. Only concerning Poland, the situation of absolute unplanned flows worsened, as the country has to incorporate more unplanned flows into its grid compared to before the policy intervention.

\vspace{0.3cm}

Figure 2 further decompose the effect of the policy intervention on absolute unplanned flows. Here, unplanned flows at the specific borders are shows. Details about the OLS regressions can be found in Table 8 in the Appendix. One can see, that very small changes occurred in connection with Hungary and mainly with Slovakia. In the lower part of the picture, more decreasing absolute unplanned flows are found: at the Austrian-German and the Czech-Slovakian interconnection. Also flows from the Czech Republic to Germany and Austria show significant reductions. Nevertheless, the OLS results suggest that there are also positive developments regarding absolute unplanned flows.

\begin{figure}[h!]
\centering
\caption{OLS results of directed absolute unplanned flows}
	\includegraphics[width=1.0\textwidth]{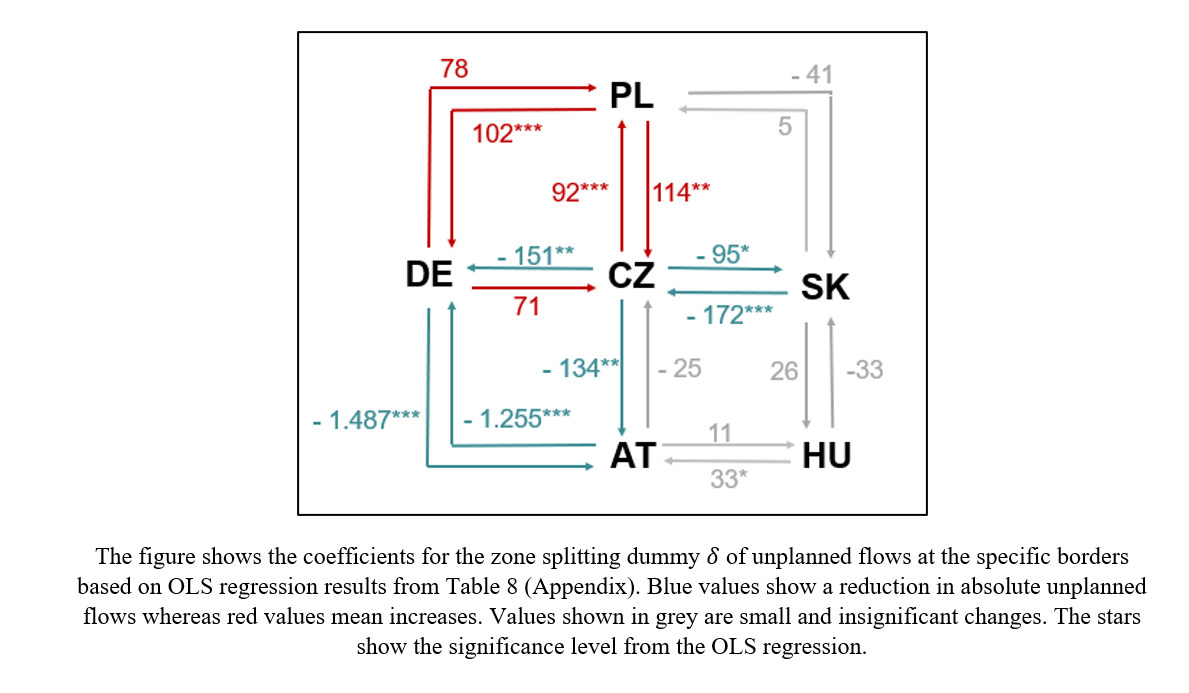}
\end{figure}

 Those mainly seem to be connected to Poland. Interestingly, the increase seems to take place in more than one direction: from Poland to the Czech Republic ($114$ MWh), and from the Czech Republic to Poland ($92$ MWh); from Poland to Germany ($102$ MWh) and also the estimated coefficient for flows from Germany to Poland is positive with $78$ MWh but not significant. Hence, the regression analysis with OLS suggests that for Poland there is still a deviation between planned flows and physical movement through the grid. Moreover, the increase in the deviation seems to be connected to Germany and the Czech Republic. 
 
 \newpage

\section{Robustness Checks}

In the baseline OLS regression, I use 107 control variables, like residual load, conventional electricity generation or the difference between forecasted and actual renewable generation values. It is well conceivable that for example wind or solar PV feed-in changes over the seasons. For example solar PV feed-in might have varying effects on the outcome variable in summer and in winter months as well as between day and night hours. Normally, one could think of much more flexible specifications of control variables by adding for example interaction effects between solar PV fluctuation and the hour of the day or the month. However, if one would use all the possible interaction effects, this would lead to a large set of over 4,000 control variables. They namely are electricity generation values for Germany, Austria, Poland, the Czech Republic, Hungary, and Slovakia (60 variables), load (6 variables) and residual load differences (8 variables). At the TSO level, the data set comprises actual and forecasted solar PV, offshore and onshore feed-in as well as the difference between forecasted and actual feed-in (36 variables). Further, residual load data for the TSOs were collected (5 variables). Residual load, residual load differences, and renewable generation are allowed to take a quadratic form as well (52 variables). Moreover, I allow interactions of all features (apart from the quadratic terms) with the month of the year (1,380 variables)  and the hour of the day (2,760 variables). \footnote{I re-estimate the equations (2) and (3) with a larger number of possible regressors $p$ stored in the vectors $\dot{X}$ and $\dot{Z}$: 	
			\begin{align*}
		\dot{planned}_{i,t} = \alpha_0 + \alpha_1 \delta_{t}+  \alpha_2 \dot{X}_{i,t} + \zeta_{i,t}  \qquad \qquad  (4)
	\end{align*}
\begin{align*}
		\dot{|unplanned|}_{i,t} = \beta_0 + \beta_1 \delta_{t}+  \beta_2 \dot{Z}_{i,t} + \xi_{i,t}\qquad \qquad  (5)
	\end{align*}
		
}
	So as a robustness check I use Post-Lasso Double Selection and Partialling-out approaches by Chernozhukov, Hansen, and Spindler (2019) to select from the large set of potential features. A good  introduction to the modern machine learning methods can also be found in Ahrens, Hansen, and Schaffer (2020). For the usage of the package `hdm` in R, I refer to Spindler, Chernozhukov, and Hansen (2016).
	\vspace{0.3cm}
	
	 If the variable selection is applied to the data set with planned flows between Germany and Austria as the outcome, 95 out of 4,200 possible features are selected with an adjusted $R^2$ of 0.80.\footnote{Here $y_i$ is regressed on the vector of all possible controls $\dot{X}$ by Post-Lasso estimation in order to select the most predictive features $s$. To control for time trends, hours, weekdays, and months are added as dummy variables. The split coefficient delta is not included into the set of potential controls.} 7 of the selected features are conventional production values from Germany and Austria, The largest fraction of the selected features is made up by interactions of solar PV and wind generation from Germany and conventional generation values from CZ/PL/SK/HU with months. Also, some German and Austrian conventional generation variables and their interactions with hours are chosen. Further, onshore and offshore wind feed-in for the 50Hertz and Tennet regions are selected. Renewable generation from the other countries plays a minor role as well as residual load and load differences. Further, 5 non-interacted renewable generation variables in linear and 7 in quadratic form are chosen by the approach. One can also find 19 interactions with hours of the day . The biggest part of them is made up by interactions of hours with the difference between Austrian forecasted and actual PV generation (11 interactions). That is not what one would expect, as planned flows should be rather driven by forecasted generation only. 35 selected features are moreover interactions of various variables with months. If instead Polish planned flows are used as the dependent variable, 124 features are selected. The adjusted $R^2$ is 0.421, which is higher than the one found for the OLS regression in Table 3. Again, the largest part is made up by interactions with hours (57) and months (46). Especially, coal generation from Germany and actual and forecasted offshore and onshore wind production from Tennet and 50Hertz are taken into account. Once more, it seems not so plausible, that here actual instead of forecasted generation is selected. 
	 
	 \vspace{0.3cm}
	 
	In Table 6, results from the baseline OLS regression are shown together with results from the Double Selection and Partialling-out approaches are presented with planned flows between Germany and Austria, for the Czech Republic, Poland, Slovakia, and Hungary as the dependent variables.

\begin{table}[h!]
	\begin{center}
		\scalebox{0.8}{
			\begin{tabular}{ m{3cm} m{3cm} m{3cm} m{3cm}  m{3cm} m{3cm}} 
				\hline
				\hline
				\multicolumn{6}{c}{\textbf{Dependent variable: Planned flows} } \\ 
				\hline \\[-1.8ex] 
				& DE-AT & CZ & PL & SK & HU \\ 
				& (1) & (2) & (3) & (4) & (5) \\ 
				\hline \\[-1.8ex] 
				
				OLS & $-$2.914$^{***}$ & 0.139  &  0.350$^{***}$ & 0.073 & 0.281$^{***}$ \\  
				& (0.230) & (0.173) & (0.056) &  (0.119) &  (0.060) \\ 
				\hline
				Adjusted $R^2$ & 0.811 & 0.305 & 0.353 & 0.403 & 0.452 \\
				\hline \\[-1.8ex] 
				Double & $-$3.865$^{***}$ & 0.031 & 0.073$^{***}$  & $-$0.085$^{.}$ & 0.171$^{***}$ \\ 
				Selection & (0.108) & (0.062) & (0.019) & (0.046) &  (0.023)  \\ 
				&   \\ 
				\hline \\[-1.8ex] 
				Partialling & $-$1.447$^{***}$ & $-$0.012 & 0.116$^{***}$ & $-$0.138$^{***}$ & 0.097$^{***}$ \\ 
				out  & (0.067) & (0.049) & (0.022) & (0.035) &  (0.020)  \\
				&\\
				\hline
				Adjusted $R^2$ & 0.804 & 0.331 & 0.421 & 0.458 & 0.495 \\
				\hline
				&   \\  
				\textit{Note:} &  \multicolumn{5}{r}{ Signif. codes for OLS: $^{*}$p$<$0.1; $^{**}$p$<$0.05; $^{***}$p$<$0.01}\\
				& \multicolumn{5}{r}{ Signif. codes for Double Selection and Partialling out: $^{***}$ 0.001 $^{**}$  0.01 $^{*}$ 0.05 } \\ 
				\hline
				\hline
		\end{tabular}}
	\end{center}
	\caption{Regression results OLS, Double Selection, Partialling out - planned flows}
\end{table}

For planned flows between Germany and Austria, the direction and significance level of the treatment effect found by OLS are confirmed. Post Lasso Double Selection yields a more negative and Partialling-out a less negative coefficient. As well, for Poland and Hungary, the significance and direction of the OLS treatment effect are approved. Still, for both bidding zones, the modern methods find smaller effect sizes. The baseline OLS estimation results in insignificant estimates for the Czech Republic and Slovakia. For the former, this finding is confirmed. Here, the sign of the coefficient flips for Partialling out but stays relatively close to zero and insignificant. Concerning Slovakia, Double Selection yields a negative but insignificant effect. Partialling out as well finds a negative effect, but it becomes significant. Hence, for Slovakia, the size and sign of the effect of the bidding zone split remain rather unclear.

\vspace{0.3cm}

Regarding unplanned flows, table 7 shows, which variables are chosen by the approach, when I use absolute unplanned flows between Germany and Austria (column 2) and Poland (column 3) as the dependent variable. In column 2, conventional electricity production seems to play a larger role. Renewable electricity generation instead is found when having a closer look at the interaction effects. For instance, the approach selects 15 interactions effects of German wind and solar PV generation with the month of the year. Overall, 77 variables are chosen. If Polish absolute unplanned flows is the dependent variable, 94 features are selected. Again, the main part is made up by interactions with the month of the year. 18 of them are made up by German and Austrian wind and solar PV interactions.  	
\begin{table}[h!]
	\begin{center}
		\scalebox{0.9}{
			\begin{tabular}{ m{6cm} m{4cm} m{3cm} m{3cm} } 
				\hline
				\hline
				Category & Number of  &  DE-AT & PL   \\ 
				&	potential features & & \\
				& (1) & (2) & (3)\\
				\hline \\[-1.8ex] 
				Wind/PV German TSOs/APG & 36 & 1 & 3 \\
				Other gen. DE/AT & 22  & 9 & 6 \\
				Wind/PV PL/CZ/SK/HU & 4 & 0 & 1 \\
				Other gen. PL/CZ/SK/HU & 34 & 8 & 10 \\
				(Residual) load (differences) & 19 & 2 & 1 \\
				Quadratic terms & 52 & 2 & 3 \\
				planned flows DE-AT & 2 & - & 2 \\
				Interactions with hour & 1,380 & 13 & 21 \\
				Interactions with month & 2,760 & 42 & 47 \\
				\hline
				\textbf{sum} & \textbf{4,200} & \textbf{77} & \textbf{94}\\					
				\hline
			   \multicolumn{4}{l}{\textit{Note:} The table shows the number of potential control variables which are chosen by the } \\
				\multicolumn{4}{l}{Double selection approach when I use absolute unplanned flows between Germany and Austria } \\
				\multicolumn{4}{l}{(column 2) or Poland (column 3) as the outcome. (The 28 quadratic terms are not interacted} \\
				\multicolumn{4}{l}{with hour or month.)} \\
				\hline
				\hline
		\end{tabular}}
	\end{center}
	\caption{Examples of variable selection Double Selection Approach}
\end{table}

 Further, table 8 shows regression results for the basline OLS, Double Selection and Partialling-out with absolute unplanned flows as the outcomes. 
 \begin{table}[h!]
 	\begin{center}
 		\scalebox{0.8}{
 			\begin{tabular}{ m{3cm} m{3cm} m{3cm} m{3cm}  m{3cm} m{3cm}} 
 				\hline
 				\hline
 				\multicolumn{6}{c}{\textbf{Dependent variable: Absolute unplanned flows} } \\ 
 				\hline \\[-1.8ex] 
 				& DE-AT & CZ & PL & SK & HU \\ 
 				& (1) & (2) & (3) & (4) & (5) \\ 
 				\hline \\[-1.8ex] 
 				
 				OLS & $-$2.696$^{***}$ & $-$0.408$^{***}$  &  0.312$^{**}$ & $-$0.314$^{***}$ & 0.025 \\  
 				& (0.254) & (0.147) & (0.132) &  (0.092) &  (0.052) \\ 
 				&\\
 				\hline \\[-1.8ex] 
 				Double & $-$3.459$^{***}$ & $-$0.579$^{***}$ & -0.072  & $-$0.437$^{***}$ & $-$0.038  \\ 
 				Selection & (0.071) & (0.063) & (0.045) & (0.04) &  (0.025)  \\ 
 				&   \\ 
 				\hline \\[-1.8ex] 
 				Partialling & $-$2.748$^{***}$ & $-$0.495$^{***}$ & 0.039 & $-$0.255$^{***}$ & $-$0.008  \\ 
 				out  & (0.065) & (0.054) & (0.037) & (0.033) &  (0.021)  \\
 				&\\
 				\hline
 				&   \\  
 				\textit{Note:} &  \multicolumn{5}{r}{ Signif. codes for OLS: $^{*}$p$<$0.1; $^{**}$p$<$0.05; $^{***}$p$<$0.01}\\
 				& \multicolumn{5}{r}{ Signif. codes for Double Selection and Partialling out: $^{***}$ 0.001 $^{**}$  0.01 $^{*}$ 0.05} \\ 
 				\hline
 				\hline
 		\end{tabular}}
 	\end{center}
 	\caption{Regression results OLS, Double Selection, Partialling out - absolute unplanned flows}
 \end{table}
 
 Both machine learning methods confirm the directions and significance levels of the results found by OLS for Germany-Austria, the Czech Republic, and Slovakia. For the German-Austrian border, the Double Selection method finds a more negative estimate ($-3.459$ MWh). Applying Post-Lasso Partialling out does not alter the OLS result a lot. For the Czech Republic, the estimated treatment effects becomes more negative for Partialling out ($-0.495$ MWh) and even more for the Double Selection approach ($-0.597$ MWh). For Slovakia, Partialling out estimates a smaller ($0.255$ MWh) and Double Selection a larger ($0.437$ MWh) negative treatment effect. For Hungary, the estimates become slightly negative, but are still close to zero. As well, the estimated coefficients stay insignificant. Concerning Poland, the found effect is still positive but much smaller and insignificant when Partialling out is applied. Using the Double Selection approach, the estimated coefficient becomes even negative but also insignificant. Thus, except for Poland, the results are confirmed by the machine learning approach. For Poland, thus the effect of the bidding zone split on absolute unplanned flows remains the least clear.

\section{Conclusion}

The split of the German-Austrian bidding zone gives a unique opportunity to observe the effect of a trading regime change on unplanned cross-border flows. The joint Austrian-German bidding zone was split in October 2018. In this analysis, I exploit the effect of this policy intervention on cross-border flows between Germany, Austria, the Czech Republic, Poland, Slovakia, and Hungary. I show that planned (or scheduled) cross-border flows decreased by nearly $3$ GWh at the German-Austrian border. For the other countries, the policy intervention was leading to increasing planned flows, where the strongest effect can be found for Poland, followed by Hungary. For absolute unplanned flows, the treatment effects are more diverse. Indeed, absolute unplanned flows between Germany and Austria substantially declined by $2.7$ GWh. Moreover, the situation for the Czech Republic and Slovakia improved. Here, a reduction in absolute unplanned flows of $40\%$ respectively $45\%$ compared to the pre-treatment mean is found after the policy intervention. The results are significant for all observed units but for Hungary, where the effect size is close to zero.  

\vspace{0.3cm}

Nevertheless, it seems like Poland is still suffering from a high amount of absolute unplanned flows. Further, it is unclear, whether the situation for this country even worsened after the split of the joint bidding zone. I find a large and significant increase in Polish unplanned flows after the policy intervention by applying OLS with a large set of control variables ($+312$ MWh). Using an even larger set of potential controls and applying machine learning techniques, yields smaller and in one case even insignificant effects. Further decomposing flows that are related to Poland does also not show a clear cause of increasing absolute unplanned flows. Thus, with regard to Poland, further analyzes are needed. It would also be interesting to study, whether a further split up of Germany into two price zones would resolve the situation to understand the causes of the lacking decrease (or probably even increase) in absolute unplanned flows for Poland which should be part of the political concern and future research.

\vspace{0.3cm}

 This analysis gives insights into the effect of the split of a large bidding zone on planned and absolute unplanned cross-border flows. In future research, it would be interesting to also look at the financial implications of the split of the German Austrian bidding zone. It may be that through the policy intervention, the need for re-dispatch actions declined with the reduction of absolute unplanned flows. This decline can be analyzed and expressed monetarily.

\newpage

\section{References}

ACER (2015), ``Opinion 09-2015 on the compliance of NRA's decision approving methods of cross-border capacity allocation in the CEE Region'', available online: 
\url{https://documents.acer.europa.eu/Official_documents/Acts_of_the_Agency/Opinions/Opinions/ACER%20Opinion%2009-2015.pdf}, accessed on 03/21/2021.

\vspace{0.1cm}

ACER (2016), ``Decision of the Agency for the Cooperation of Energy Regulators No. 06/2016 on the electricity transmission system operators' proposal for the determination of capacity calculation regions'', available online: \url{https://documents.acer.europa.eu/Official_documents/Acts_of_the_Agency/Individual%20decisions/ACER%20Decision%2006-2016%20on%20CCR.pdf}, accessed on 03/24/2021.

\vspace{0.1cm}

ACER (2018), `` Methodological Paper: Unscheduled flows'', available online:
\url{https://extranet.acer.europa.eu/en/Electricity/Market%20monitoring/Documents_Public/ACER%20Methodological%20paper%20-%20Unscheduled%20flows.pdf} , accessed on 06/01/2021.

\vspace{0.1cm}

ACER (2019), ``Market Monitoring Report 2018 - Electricity Wholesale Markets Volume'', available online: \url{https://extranet.acer.europa.eu/Official_documents/Acts_of_the_Agency/Publication/ACER%20Market%20Monitoring%20Report%202018%20-%20Electricity%20Wholesale%20Markets%20Volume.pdf}, accessed on 08/01/2021.

\vspace{0.1cm}

ACER (2021), ``The Agency'', available online: \url{http://acer.europa.eu/the-agency/about-acer}, accessed on 03/23/2022.
\vspace{0.1cm}

Antweiler, W. (2016), ``Cross-border trade in electricity'', \textit{Journal of International Economics} \textbf{101}, 42-51.
\vspace{0.1cm}

Ahrens, A., Hansen, C., and Schaffer, M. (2020), ``lassopack: Model selection and prediction with regularized regression in Stata'', \textit{The Stata Journal} \textbf{20}, 176-235.
\vspace{0.1cm}

APG (2021), ``End of the German-Austrian Electriciy Price Zone - What does that mean?'', available online: \url{file:///D:/Forschung/Bidding%20Zone%20Split/PDF_Downloads_Wissen_STROMPREISZONE_EN.pdf}, accessed on 05/07/2021.
\vspace{0.1cm}

Belloni, A., Chernozhukov, V., and Hansen, C. (2014), ``Inference on treatment effects after selection among high-dimensional controls'', \textit{Review of Economic Studies} \textbf{81}, 608-650.
\vspace{0.1cm}

Bem\v{s}, J., Kr\'{a}l\'{i}k, T., and Kn\'{a}pek, J. (2016), ``Bidding Zones Reconfiguration - Current Issues. Literature Review, Criterions and Social Welfare'', \textit{2016 2nd International Conference on Intelligent Green Building and Smart Grid}, 1-6.
\vspace{0.1cm}

Borowski, P. (2020), ``Zonal and nodal Models of Energy Markets in Euopean Union'', \textit{Energies} \textbf{13, 4182}.

\vspace{0.1cm}

Bundesnetzagentur (2022), ``Engpassbewirtschaftung an der Grenze zu \"Osterreich'', available online: \url{https://www.smard.de/page/home/wiki-article/446/204212}, accessed on 01/28/2022.

\vspace{0.1cm}

\v{C}EPS, MAVIR, PSE, and SEPS (2013), ``Unplanned flows in the CEE region in relation to the common market area Germany - Austria'', \textit{Joint study}.

\vspace{0.1cm}

\v{C}EPS, MAVIR, PSE, and SEPS (2017), ``V4 Transmission system operators (v{C}EPS, MAVIR, PSE, SEPS) and TRANSELECTRICA are deeply concerned by nontransparent developments around the implementation of congestion management at the Austrian-German border'', \textit{Press release}.

\vspace{0.1cm}

Chernozhukov, V., Hansen, C., Spindler, M. (2019), ``High-Dimensional Metrics in R'', ArXiv Working Paper No. arXiv:1603.01700. https://arxiv.org/abs/1603.01700.

\vspace{0.1cm}

Cl\'{o}, S., Cataldi, A.,  and Zoppoli, P. (2015). ``The merit-oder effect on the Italien power market: The impact of solar and wind generation on national wholesale electricity prices'', \textit{Energy Policy 77}, p.79-88.
\vspace{0.1cm}

ENTSO-E (2014), ``Technical Report. Bidding Zones Review Process'', available online: \url{https://extranet.acer.europa.eu/en/Electricity/MARKET-CODES/CAPACITY-ALLOCATION-AND-CONGESTION-MANAGEMENT/17%20BZR/Action%201b%20-%20ENTSO-E%20Technical%20Report%2c%20Bidding%20Zones%20Review%20Process.pdf}, accessed on 04/01/2021.

\vspace{0.1cm}

ENTSO-E Transparency Platform (2021): ``Total scheduled commercial exchanges from explicit and implicit allocations''. Available online:
\url{https://transparency.entsoe.eu/transmission-domain/r2/scheduledCommercialExchangesDayAhead/show.html}, accessed on 01/28/2022.
\vspace{0.1cm}

Frontier Economics, and Consentec (2011), ``Relevance of established national bidding areas for European power market integration - an approach to welfare oriented evaluation'', \textit{Executive Summary for the Bundesnetzagentur}.
\vspace{0.1cm}

Gugler, K., Haxhimusa, A., and Liebensteiner, M. (2018), ``Integration of European Electricity Markets: Evidence from Spot Prices'', \textit{The Energy Journal} \textbf{39}, 41-66.

\vspace{0.1cm}

Guo, B., Newbery, D., and Gissey, G. (2019), ``The impact of unilateral carbon taxes on cross-border electricity trading'', \textit{Cambridge Working Papers in Economics} \textbf{1951}.

\vspace{0.1cm}
 J\'{o}nsson, T., Pinson, P., and Madsen, H. (2010), ``On the market impact of wind energy forecasts'', \textit{Energy Economics} \textbf{32}, 313-320.

\vspace{0.1cm}
Keppler, J., Phan, S., and Le Pen, Y. (2016), ``The Impacts of Variable Renewable Production and Market Coupling on the Convergence of French and German Electricity Prices'', \textit{The Energy Journal} \textbf{37}, 343-360.

\vspace{0.1cm}
Moran, M., Shapiro, H., Boettner, D., and Bailey, M. (2011), ``Fundamentals of engineering thermodynamics", seventh edition, John Wiley, Hoboken.

\vspace{0.1cm}
Newey, W., and West, K. (1994), ``Automatic Lag Selection in Covariance Matrix Estimation'', \textit{Review of Economic Studies} \textbf{61}, 631-653.
\vspace{0.1cm}

Singh, A., Frei, T., Chokani, N., and Abhari, R. (2016), ``Impact of unplanned power flows in interconnected transmission systems - Case study of Central Eastern European region'', \textit{Energy Policy} \textbf{91}, 287-303.
\vspace{0.1cm}

Sk\r{a}nlund, A. Schemde, A., Tennbakk, B, Gravdehaug, G., and Grondahl, R. (2013), ``Loop flows - Final advice'', \textit{THEMA Report prepared for the European Comission} \textbf{2013-36}.
\vspace{0.1cm}

Spindler, M., Chernozhukov, V., and Hansen, C. (2016), ``High-Dimensional Metrics '', \textit{R-Package: https://cran.r-project.org/web/packages/hdm/index.html.}.
\vspace{0.1cm}

Taniguchi, S. (2020), ``Examining the causality structures of electricity interchange and variable renewable energy: a comparison between Japan and Denmark'', \textit{Asia-Pacific Journal of Regional Science} \textbf{4}, 159-191.
\vspace{0.1cm}

Wasserbacher, H., and Spindler, M. (2021), ``Machine Learning for financial forecasting, planning and analysis: recent developments and pitfalls'', \textit{Digital Finance} \textbf{4}, 63-88.

\vspace{0.1cm}

Zugno, M., Pinson, P., and Madsen, H. (2013), ``Impact of Wind Power Generation on European Cross-Border Power Flows'', \textit{IEEE Transaction on Power Systems} \textbf{28}, 3566-3575.

\newpage
\appendix

\section{Appendix A: Additional figures}

\begin{figure}[h!]
\centering
\caption{Unplanned cross-border flows}
	\fbox{\includegraphics[width=1.0\textwidth]{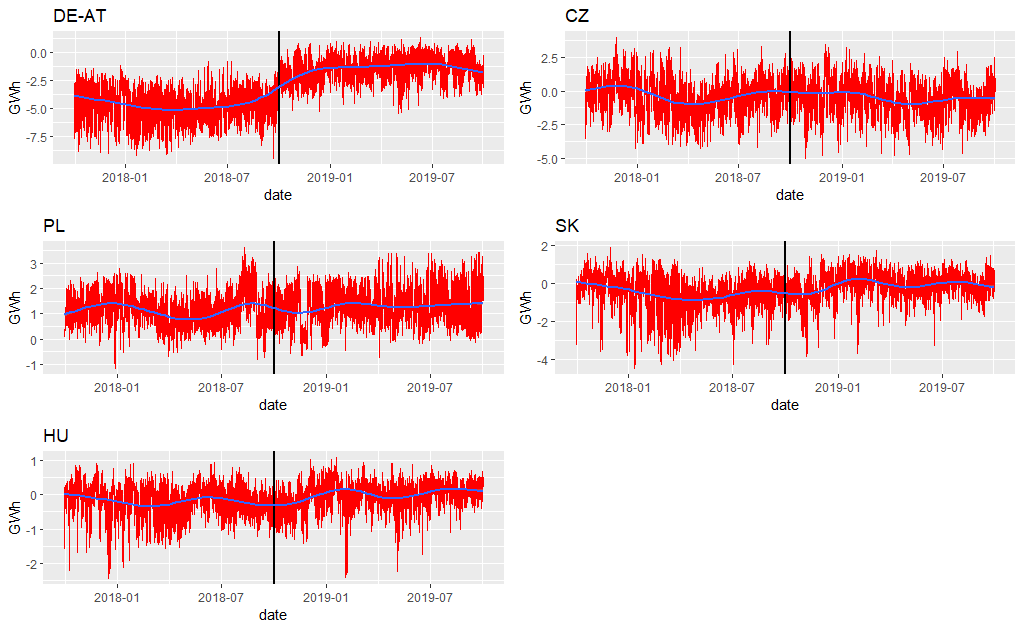}}
\end{figure}

The plots in Figure 3 display unplanned cross-border flows between Germany and Austria and for Poland, the Czech Republic, Slovakia, and Hungary between October 2017 and October 2019. The black vertical line marks the policy intervention. The horizontal curved line again is a smoother function. Unplanned flows are calculated as the difference of the sums of physical minus the sums of planned flows for the respective unit. 

\vspace{0.3cm}

During peak hours, the electricity demand generally is higher than during off-peak hours.\footnote{On-peak hours are often defined between 8 A. M. and 8 P. M. on weekdays, whereas off-peak hours are the night hours between 8 P. M. and 8 A. M. on weekdays and all hours on weekends (Moran et al. 2011).} Figure 4 gives an overview of how absolute unplanned flows vary over the day and on different weekdays. For every hour of the day, therefore one average value over the whole sample (pre- and post-treatment) is calculated and plotted for each day of the week. One can see, that average absolute unplanned flows are on the highest levels in the Czech Republic and on the German-Austrian border. The development is quite stable in the Czech Republic throughout the day as well as between weekdays. One can not observe great outliers. Overall it still seems like unplanned flows are more stable in peak hours and slightly increase and vary more in off-peak hours. At the German-Austrian border, unplanned flows are on a much lower level on the weekend. Moreover, absolute unplanned flows reach a plunge in the early morning hours. They then increase on weekdays, stay on a quite stable level for the peak hours and realize the highest levels between 4 and 8 P. M. In Hungary, absolute unplanned flows are low and quite evenly distributed throughout the day. Also, the weekday does not seem to have a big influence of the amount of unplanned flows. A small peak can be observed in the early morning hours. The same holds for Slovakia, nevertheless here flows are on a higher level and the peak in the morning hours is more pronounced than in Hungary. For Poland again unplanned flows are lowest on the weekend. Also one can see that again the development throughout the day is quite stable and on a lower level within peak hours. Overall, apart from the German-Austrian interconnection, absolute unplanned flows descriptively seem to be somewhat higher and less stable in off-peak hours, while they are quite steady and on a lower level in peak hours.

\begin{figure}[h!]
\centering
\caption{Absolute unplanned hourly cross-border flows per weekday}
	\fbox{\includegraphics[width=1.0\textwidth]{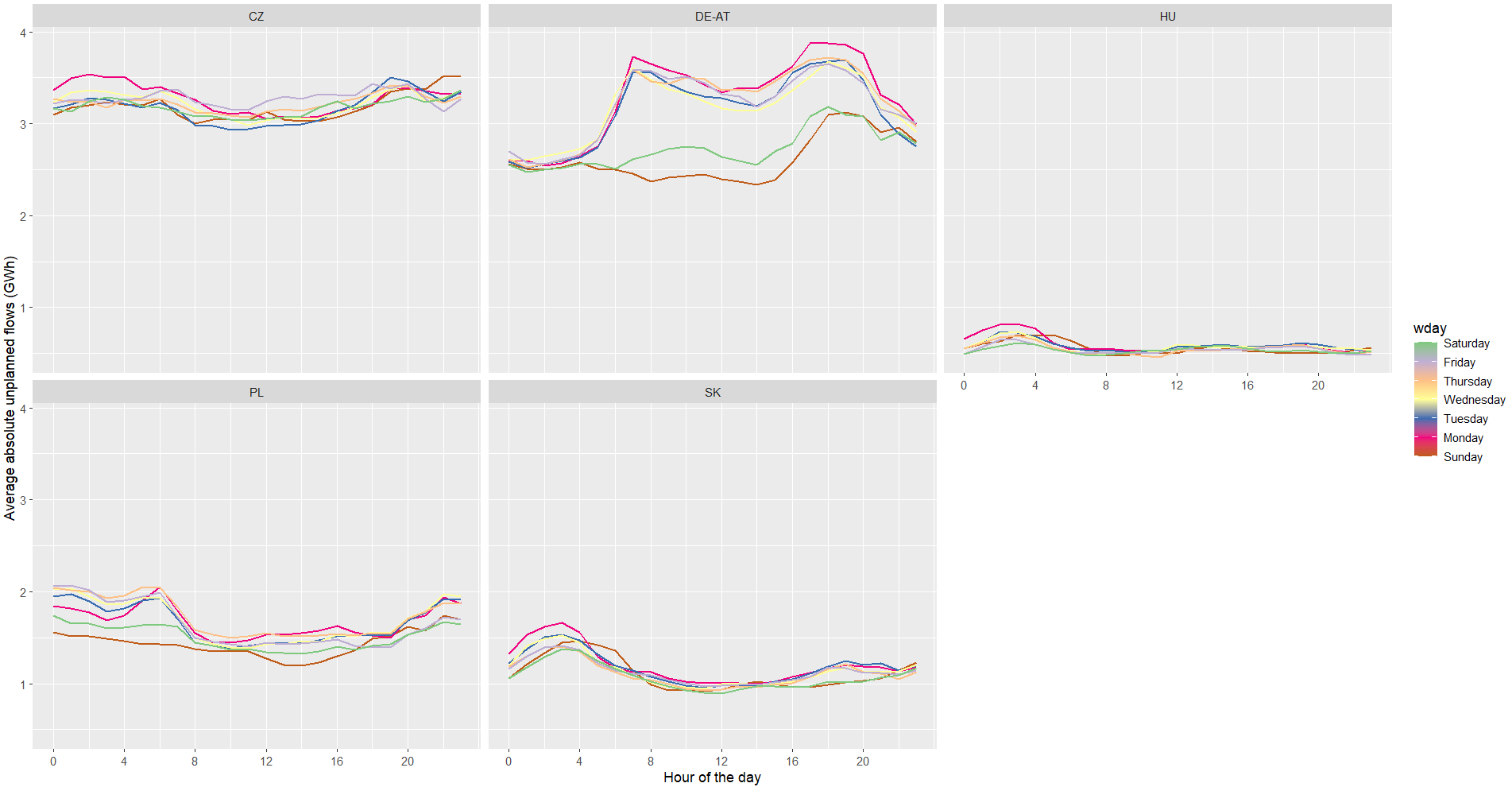}}
\end{figure}

\newpage
\section{Appendix B: Additional tables}

Table 9 gives an overview of the regression of physical flows as dependent variable. The rest of the regression specification is as described in equation (3?).

\begin{table}[h!]
\begin{center}
\scalebox{0.8}{
\begin{tabular}{ m{3cm} m{3cm} m{3cm} m{3cm} m{3cm} m{3cm}} 
\hline
\hline
  \multicolumn{6}{c}{\textbf{Dependent variable: Physical flows} } \\ 
\hline \\[-1.8ex] 
 & DE-AT & CZ & PL & SK & HU \\ 
& (1) & (2) & (3) & (4) & (5) \\
\hline \\[-1.8ex] 
delta & $-$0.126 & 0.088 & 0.280$^{**}$ & 0.284$^{**}$ & 0.218$^{***}$  \\ 
  & (0.081) & (0.151) & (0.137) & (0.138) & (0.074)  \\ 
  &   \\  
\hline \\
Observations & 17,518 & 17,518 & 17,518 & 17,518 & 17,518 \\ 
Adjusted R$^{2}$ & 0.829 & 0.693 & 0.482 & 0.727 & 0.712 \\ 
\hline
\textit{Note:}  & \multicolumn{4}{r}{$^{*}$p$<$0.1; $^{**}$p$<$0.05; $^{***}$p$<$0.01} \\ 
 \multicolumn{6}{l}{Standard errors are HAC robust. Lags are chosen with the Newey-West (1994) method.} \\
\hline
 \hline
\end{tabular}}
\end{center}
\caption{Regression results physical flows}
\end{table}

For the estimation, physical flows that represent the movement of electricity flows through the grid, are the dependent variables. The rest of the regression specification stays as in equation (2). Only instead of forecasted renewable generation for the German and Austrian TSOs, actual production values are taken into account as control variables. One can observe that physical flows between Germany and Austria decreased but to a much lesser extent than planned flows. The estimated coefficient is insignificant and $-126$ MWh for that border. Here, the sign is contrary to the one presented in Table 1, where sole averages are computed. For the Czech Republic, physical flows slightly increase after the policy intervention ($88 MWh$). Still, also here the estimated coefficient is insignificant. Larger and significant increases in physical flows are estimated for Slovakia ($284$ MWh), Poland ($280$ MWh), and Hungary ($218$ MWh).

\newpage
In Table 10, absolute unplanned flows are further separated into directed flows per interconnection.

\begin{table}[h!]
	\begin{center}
		\scalebox{0.8}{
			\begin{tabular}{ m{3cm} m{3cm} m{3cm} m{3cm}  m{3cm} } 
				\hline
				\hline
				\multicolumn{5}{c}{\textbf{Dependent variable: Directed absolute unplanned flows per border}} \\ 
				\hline \\[-1.8ex] 
				& CZ-DE &  DE-CZ & CZ-AT & AT-CZ   \\ 
				\hline \\[-1.8ex] 
				delta & -0.113 & 0.035 & $-$0.191$^{***}$ & $-$0.043  \\ 
				& (0.074) & (0.068) & (0.060) & (0.028)   \\ 
				\hline \\
				Adjusted R$^{2}$ & 0.411 & 0.192 & 0.761 & 0.636 \\ 
				\hline
				\hline
				& \\
				& PL-CZ &  CZ-PL &  PL-DE  & DE-PL \\
				\hline \\[-1.8ex] 
				delta & 0.138$^{***}$ & 0.063$^{**}$ & 0.073$^{***}$ & 0.083 \\
				& (0.049) & (0.027) & (0.027) & (0.068)\\
				\hline \\
				Adjusted R$^{2}$ & 0.490 & 0.565 & 0.519 & 0.521 \\ 
				\hline
				\hline
				& \\
				& PL-SK &  SK-PL &  CZ-SK  & SK-CZ \\
				\hline \\[-1.8ex] 
				delta & $-$0.044 & $-$0.0001 & $-$0.099$^{**}$ & $-$0.198$^{***}$ \\
				& (0.037) & (0.005) & (0.047) & (0.048)\\
				\hline \\
				Adjusted R$^{2}$ & 0.517 & 0.193 & 0.391 & 0.423 \\
				\hline
				\hline
				& \\
				& SK-HU &  HU-SK &  AT-HU  & HU-AT \\
				\hline \\[-1.8ex] 
				delta & 0.028 & $-$0.042 & 0.002 & 0.039$^{**}$ \\
				& (0.032) & (0.027) & (0.028) & (0.019)\\
				\hline \\
				Adjusted R$^{2}$ & 0.147 & 0.343 & 0.234 & 0.372 \\
				\hline
				\hline 
				\textit{Note:}  & \multicolumn{4}{r}{$^{*}$p$<$0.1; $^{**}$p$<$0.05; $^{***}$p$<$0.01} \\ 
				\multicolumn{5}{l}{Standard errors are HAC robust. Lags are chosen with the Newey-West (1994) method.} \\
				\hline
				\hline
		\end{tabular}}
	\end{center}
	\caption{Regression results absolute unplanned flows per border}
\end{table}

\newpage

To analyze the effect of peak hours, I re-estimate equation (3) but add a dummy for peak hours \textit{$peak_t$} that is equal to 1 for on-peak and equal to 0 for off-peak hours. Also, I interact the split dummy $\delta_t$ with \textit{$peak_t$}:

\begin{align*}
unplanned_{i,t} = \theta_0 + \theta_1 \delta_{t} +  \theta_2 peak_t + \theta_3 \delta_t \times peak_t + \theta_4 X_{i,t} +  \pi_t + \nu_{i,t}\qquad \qquad  (5)
\end{align*}

With $\pi_t$, only fixed effects for the month of the year and the weekday are added. With that specification it is possible to test, whether absolute unplanned flows depend on peak hours ($\theta_2 \neq 0$) and whether the effect of the split coefficient depends on peak hours ($\theta_3 \neq 0$). The results are presented by table 11.

\begin{table}[h!]
\begin{center}
\scalebox{0.7}{ 
\begin{tabular}{  m{3cm}   m{3cm}  m{3cm}   m{3cm}  m{3cm}  m{3cm}  }
\\[-1.8ex]\hline 
\hline \\[-1.8ex] 
& \multicolumn{5}{c}{\textbf{Dependent variable: Absolute unplanned flows per border}} \\ 
\hline
\\[-1.8ex]
 & DE-AT & CZ & PL & SK &  HU  \\ 
\\ & (1) & (2) & (3) & (4) & (5)\\ 
\hline \\[-1.8ex] 
peak & 0.121$^{*}$ & $-$0.009 & $-$0.268$^{***}$ & 0.010 & 0.042$^{**}$ \\ 
  & (0.070) & (0.051) & (0.035) & (0.029) & (0.020) \\ 
  &  \\ 
 delta & $-$2.656$^{***}$ & $-$0.301$^{**}$ & 0.248$^{**}$ & $-$0.281$^{***}$ & 0.044 \\ 
  & (0.221) & (0.144) & (0.122) & (0.092) & (0.052) \\ 
  & \\ 
 peak:delta & $-$0.169$^{**}$ & 0.068 & 0.231$^{***}$ & 0.094$^{***}$ & 0.040$^{*}$ \\ 
  & (0.082) & (0.061) & (0.044) & (0.036) & (0.022) \\ 
  &  \\ 
\hline \\[-1.8ex] 
Observations & 17,518 &  17,518 &   17,518 &   17,518 &   17,518  \\ 
Adjusted R$^{2}$ & 0.786 &  0.329  & 0.515  & 0.383 &   0.306  \\
\hline 
\hline \\[-1.8ex] 
\textit{Note:}  & \multicolumn{5}{r}{$^{*}$p$<$0.1; $^{**}$p$<$0.05; $^{***}$p$<$0.01} \\ 
 \multicolumn{6}{r}{Standard errors are HAC robust. Lags are chosen with the Newey-West (1994) method.} \\
\hline
\end{tabular}}
\end{center}
\caption{Regression results peak hours}
\end{table} 

\textit{Peak} shows the estimated effects of on-peak hours before the treatment. For the Czech Republic and Slovakia, the found coefficients are close to zero. The estimated coefficient for Poland is strongly negative and significant. Here, the largest effect is found which means that absolute unplanned flows are lower by $268$ MWh in peak hours compared to off-peak hours before the treatment. For flows between Germany and Austria, the effect size is positive and significant at the $10\%$ level. Here, absolute unplanned flows are $121$ MWh higher in pre-treatment peak hours compared to pre-treatment off-peak hours. For Hungary, the effect is also positive but much smaller. \textit{delta} gives the estimated effect for post-treatment off-peak hours. The largest coefficient is found for DE-AT ($-2.66$ GWh). I further find negative effects for the Czech Republic and Slovakia. For Poland and Hungary, it is positive. The interaction term \textit{delta:peak} allows the population effect of post-treatment values on absolute unplanned flows to depend on on-peak hours. For all columns I find a value that is not $0$. The differential between on-peak post-treatment hours relative to off-peak pre-treatment hours is positive for Poland ($211$ MWh), and Hungary ($126$ MWh). For flows between Germany and Austria and for the Czech Republic and Slovakia, it is negative ($-2.704$ GWh; $-242$ MWh; $-177$ MWh).\footnote{The differential between on-peak post-treatment hours and off-peak pre-treatment hours is calculated as the sum of $\hat{\theta}_1$, $\hat{\theta}_2$ and $\hat{\theta}_3$.}

\vspace{0.3cm}

\v{C}EPS et al. (2013) and ACER (2015) argue, that unplanned flows in the region mainly occur when scheduled flows from Germany to Austria exceed $3,000$ MWh. Thus, the data is filtered for hours where commercially traded flows on the respective border exceed this threshold. This is the case for 11,898 of the 17,518 observed hours, of which 8,045 are pre-treatment and 3,853 are post-treatment hours. Thus, in around $92\%$ of pre-treatment hours scheduled flows between DE-AT exceeded the threshold whereas this is the case for only $44\%$ of post-treatment hours. So, the occurrence of those high planned flows between Austria and Germany substantially declined. The regression results are shown in Table 12.

\begin{table}[h!]
\begin{center}
\scalebox{0.7}{
 \begin{tabular}{  m{3cm}   m{3cm}  m{3cm}   m{3cm}  m{3cm}  m{3cm}  } 
\hline
\hline
 \multicolumn{6}{c}{\textbf{Dependent variable: Absolute unplanned flows per border}} \\ 
\hline \\[-1.8ex] 
& DE-AT & CZ & PL & SK  & HU \\
& (1) & (2) & (3) & (4) & (5) \\ 
\hline
&\\
 All  & $-$2.733$^{***}$ & $-$0.301$^{**}$ &  0.349$^{***}$ & $-$0.277$^{***}$ & 0.038 \\  
& (0.232) & (0.144) & (0.132) &  (0.093) &  (0.051) \\ 
&\\
\hline
Observations & 17,518 & 17,518 & 17,518 & 17,518 & 17,518 \\
Adjusted R$^{2}$ & 0.788 & 0.335 & 0.519 & 0.385 & 0.310 \\
& \\ 
\hline
&\\
 Filtered  & $-$1.820$^{***}$ & $-$0.365$^{***}$ &  0.354$^{***}$ & $-$0.280$^{***}$ & 0.046 \\  
& (0.194) & (0.148) & (0.129) &  (0.096) &  (0.056) \\ 
&\\
\hline
Observations & 11,898 & 11,898 & 11,898 & 11,898 & 11,898 \\
Adjusted R$^{2}$ & 0.800 & 0.426 & 0.564 & 0.419 & 0.316 \\
& \\ 
\hline
\textit{Note:}  & \multicolumn{5}{r}{$^{*}$p$<$0.1; $^{**}$p$<$0.05; $^{***}$p$<$0.01} \\ 
\multicolumn{6}{r}{Standard errors are HAC robust. Lags are chosen with the Newey-West (1994) method.}\\
\hline
\hline
\end{tabular}}
\end{center}
\caption{Regression results all and filtered observations}
\end{table}

If the data is filtered as described, the largest deviation in the treatment effect in comparison with the regression with all data points is found for DE-AT. Here, the estimated treatment effect is $-1.820$ or $39.2\%$ instead of $-2.733$ GWh ($-58.9\%$). For the Czech Republic, the estimator becomes more negative. For the other countries, filtering the data does not alter the results. Still, the adjusted $R^2$ goes up for all regressions with the filtered data. It seems like, more variation in absolute unplanned flows can be explained through the explanatory variables in hours with a high amount of scheduled flows between Germany and Austria.

\end{document}